# Asymptotically Optimal Assignments In Ordinal Evaluations of Proposals[1]

A. Yavuz Oruç
Department of Electrical and Computer Engineering
University of Maryland, College Park, MD 20742

Abdullah Atmaca
Department of Computer Science
Bilkent University, Ankara, Turkey

**Abstract**

In ordinal evaluations of proposals in peer review systems, a set of proposals is assigned to a fixed set of referees so as to maximize the number of pairwise comparisons of proposals under certain referee capacity and proposal subject constraints. In this paper, the following two related problems are considered: (1) Assuming that each referee has a capacity to review $k$ out of $n$ proposals, $2 \leq k \leq n$, determine the minimum number of referees needed to ensure that each pair of proposals is reviewed by at least one referee, (2) Find an assignment that meets the lower bound determined in (1). It is easy to see that one referee is both necessary and sufficient when $k = n$, and $n(n-1)/2$ referees are both necessary and sufficient when $k = 2$. We show that 6 referees are both necessary and sufficient when $k = n/2$. We further show that 11 referees are necessary and 12 are sufficient when $k = n/3$, and 18 referees are necessary and 20 referees are sufficient when $k = n/4$. A more general lower bound of $n(n-1)/k(k-1)$ referees is also given for any $k$, $2 \leq k \leq n$, and an assignment asymptotically matching this lower bound within a factor of 2 is presented. These results are not only theoretically interesting but they also provide practical methods for efficient assignments of proposals to referees.

## 1. Introduction

Challenges confronted by funding agencies in identifying high quality proposals are well documented in the literature, see, for example, [Bornmann 2007, Oruc 2006, Andersson 2006, Langfeldt 2004, Wessely 1998, Hodgson 1997, Cicchetti 1991]. When resources are limited, as is very often the case, proposals must be ranked with respect to a number of attributes such as intellectual merit, broader impact, feasibility, etc. Broadly speaking, the evaluation of proposals consists of two interrelated tasks: (a) assignment of proposals to referees, (b) ranking and selection of proposals. The ranking and selection of proposals typically rely on cardinal (quantitative) or ordinal (preference) –based comparisons [Park 1997, Cook 2006, Hochbaum 2006]. Recently, Cook et al. demonstrated that cardinal comparisons such as using average scores of proposals could be unreliable especially when referees' scores are not normalized [Cook 2005, 2006, 2007]. They suggested that quantifying the intrinsic values of proposals may be difficult, and therefore it is more practical to rely on ordinal rankings. Ordinal and cardinal strengths of preferences have also been advocated in [Malakooti 2000] as natural extensions of ordinal comparison models. A set-covering integer programming approach was introduced in [Cook 2005] to obtain as many comparisons as possible between the proposals reviewed by a fixed set of referees. In [Cook 2007] a branch and bound algorithm was introduced to minimize the number of disagreements among referees based on pairwise comparisons of proposals. More recently, [Chen 2009] presented a maximum consensus algorithm based on complete rankings of a set of proposals by a set of referees, and [Iyer 2003, Ahn 2008, Sarabando 2009] presented dominance-based ordinal ranking and selection algorithms.

[1] This research is funded in part by the Scientific and Technological Research Council of Turkey under grant No: 109M149. This work has been submitted to the IEEE for possible publication. Copyright may be transferred without notice, after which this version may no longer be accessible.

This paper is concerned with the assignment aspect of ordinal evaluations of proposals. In an ordinal ranking, limited coupling of proposals to referees divides proposals into disjoint clusters and makes it impossible to compare proposals between clusters. For example, suppose that 6 proposals are to be assigned to 3 referees under the following incidence (matching) relation with the constraint that no referee can be assigned more than 3 proposals:

>Referee 1 can review Proposals 1,2,3,4
>Referee 2 can review Proposals 2,3,4,5
>Referee 3 can review proposals 1,4,5,6

It is obvious that it is impossible to cover all 6×5/2 = 15 possible pairs of proposals under the capacity constraint of 3 proposals per referee. It is still desirable to determine which three proposals should be assigned to each referee so that the number of pairs of proposals covered between the three referees is maximized. In this example, assigning proposals 1,2,3 to referee 1, proposals 3,4,5 to referee 2, and proposals 1,5,6 to referee 3 gives a maximum of 9 pairs of proposals.

As the example illustrates, the underlying assumption of the approach described in [Cook 2005] is that both proposals and referees are fixed a priori together with an incidence relation to specify which proposals can potentially be assigned to which referees. In contrast, we consider assignment problems in this paper with only two parameters of interest: (1) the number of proposals, $n$, and (2) the capacity of each referee, $k$, $2 \leq k \leq n$, i.e., the maximum number of proposals that can be reviewed by each referee. With these two parameters, we consider two related problems: (1) determine the minimum number of referees to ensure that each pair of proposals is reviewed by at least one referee, (2) find an assignment of a set of $n$ proposals to the minimum number of referees determined so that all pairs of proposals are covered. Our interest in these problems is motivated by the fact that referees are generally selected to meet the evaluation needs of a set of proposals rather than randomly assembled together. Thus, unlike in the assignment problems considered in [Cook 2005, 2007], minimizing the number of referees is the main objective in the assignments of proposals to referees in our work. We consider the assignments of proposals to referees both with and without referee specialties. In the first case, referees may be viewed interchangeable in terms of their expertise. This assumption generally holds for those proposal evaluation processes in which a small set of proposals with identical topics is considered, or for those in which a large set of proposals is prescreened to identify a small set of proposals for a second stage of a more intense peer review. In the second case, referees with specialties are allowed. This applies to peer review panels in which experts with a multitude of evaluation (research) specialties compare proposals with a multitude of subjects.

In both cases, it is useful to derive lower and upper bounds on the number of referees. We prove that 6 referees are both necessary and sufficient when each referee is assigned one-half of all proposals. We further show that 11 referees are necessary and 12 referees are sufficient when each referee is assigned one-third of all proposals and 18 referees are necessary and 20 referees are sufficient when each referee is assigned one-fourth of all proposals. We also give a more general lower bound of $n(n-1)/k(k-1)$ referees for any referee capacity $k$, $2 \leq k \leq n$, and present an assignment, asymptotically matching this lower bound within a factor of 2. These results are not only theoretically interesting but also provide practical methods for efficient evaluations of proposals.

The rest of the paper is organized as follows. In Section 2, we derive our lower bounds. In Section 3, assignments that match these lower bounds are presented. These results are extended to distinguishable referees in Section 4. The paper is concluded in Section 5.

**2. Lower Bounds**

Let $P = \{p_1, p_2,\ldots,p_n\}$ be a set of proposals, $n \geq 2$, and let $R = \{r_1, r_2,\ldots,r_m\}$ be a set of referees. The referees in $R$ are said to cover all $n(n-1)/2$ pairs of $n$ proposals if each pair of proposals is reviewed by at least one referee in $R$. Suppose that each referee is willing to review $k$ proposals, where $k$, $2 \leq k \leq n$. Then, for all $n(n-1)/2$ pairs of proposals to be covered by the $m$ referees, the following inequality must clearly hold:

$$m\binom{k}{2} \geq \binom{n}{2}, \quad k \geq 2 \tag{1}$$

Simplifying this inequality gives the following lower bound on the number of referees:

$$m = \left\lceil \frac{n(n-1)}{k(k-1)} \right\rceil, \quad k \geq 2 \tag{2}$$

In particular, when $k = 2$, that is, when each referee reviews 2 proposals, a minimum of $n(n-1)/2$ referees is required, and when $k=n$, one referee is required. Other constraints can be derived from this inequality. Table 1 lists the capacities of referees versus minimum numbers of referees for various values of $n$. It is obvious that when $k = n$, and $n \geq 2$, one referee will also suffice, and hence $m = 1$ is always achievable. For even $n$ and $k = n/2$, the table shows that $m$ tends to 5 as $n \to \infty$. However, for $n = 4$, Eqn. (2) implies that $m = 6$. We strengthen the lower bound to 6 for other values of $n$ as follows.

| Referee Capacity | Minimum Number of Referees ($m$) | | | | | | |
|---|---|---|---|---|---|---|---|
| | Eqn. (2) | $n = 2$ | $n = 4$ | $n = 8$ | $n = 16$ | $n = 32$ | $n \to \infty$ |
| $k = n$, $n \geq 2$ | $m \geq 1$ | $k = 2, m \geq 1$ | $k = 4, m \geq 1$ | $k = 8, m \geq 1$ | $k = 16, m \geq 1$ | $k = 32, m \geq 1$ | $m \to 1$ |
| $k = n/2$, $n \geq 4$ | $m \geq \lceil 4(n-1)/(n-2) \rceil$ | N/A | $k = 2, m \geq 6$ | $k = 4, m \geq 5$ | $k = 8, m \geq 5$ | $k = 16, m \geq 5$ | $m \to 5$ |
| $k = n/3$, $n \geq 6$ | $m \geq \lceil 9(n-1)/(n-3) \rceil$ | N/A | N/A | $k = 3, m \geq 15$ | $k = 6, m \geq 11$ | $k = 12, m \geq 10$ | $m \to 10$ |
| $k = n/4$, $n \geq 8$ | $m \geq \lceil 16(n-1)/(n-4) \rceil$ | N/A | N/A | $k = 2, m \geq 28$ | $k = 4, m \geq 20$ | $k = 8, m \geq 18$ | $m \to 17$ |

Table 1: Minimum numbers of referees with specified capacities for $n$ proposals.

**Theorem 1:**

For all even $n = 2k \geq 4$, if each referee is assigned $k$ proposals, at least 6 referees are needed to cover all pairs of $n$ proposals.

**Proof**: For $n = 4$, $k = 2$, each referee is assigned two proposals, and can therefore cover only one pair. Since there are 6 pairs of proposals in all, 6 referees are clearly necessary. For any even $n \geq 6$, without loss of generality, suppose that the first 2 referees are assigned $k$ proposals as shown below with $u$ proposals shared between them, where $u$ is an integer between 0 and $k$, and the shaded areas represent the sets of proposals assigned to the two referees:

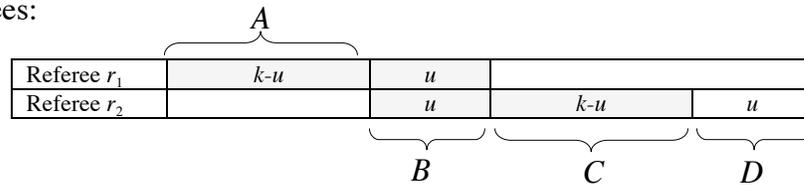

Figure 1.

Then we have the following sets of pairs of proposals that remain to be covered:

$$A \times C = \{(a,c) : a \in A, c \in C\}$$
$$A \times D = \{(a,d) : a \in A, d \in D\}$$
$$B \times D = \{(b,d) : b \in B, d \in D\} \quad (3)$$
$$C \times D = \{(c,d) : c \in C, d \in D\}$$
$$D \times D = \{(d_1,d_2) : d_1, d_2 \in D, d_1 < d_2\}$$

If $u = 0$ then $B$ and $D$ vanish, and $|A| = |C| = k$ so that the number of additional pairs of proposals that remain to be covered is given by $k^2$. Furthermore, in order to cover these $k^2$ pairs of proposals, each additional referee must be assigned at least one proposal from each of $A$ and $C$. Therefore, the number of additional referees cannot be less than

$$\left\lceil \frac{k^2}{w(k-w)} \right\rceil,$$

where $w$ denotes the number of proposals in $A$ and $k-w$ denotes the number of proposals in $C$. Since the denominator is maximized when $w = k/2$, the number of additional referees cannot be less than 4 implying that 6 referees are necessary in this case.

On the other hand, if $u = k$ then $A$ and $C$ vanish, and $|B| = |D| = k$ so that the number of additional pairs of proposals to be covered is given by $k^2 + k(k-1)/2$. But since each new referee can cover at most $k(k-1)/2$ proposals, we need at least

$$\left\lceil \frac{k^2 + k(k-1)/2}{k(k-1)/2} \right\rceil = \left\lceil \frac{3k^2 - k}{k^2 - k} \right\rceil \geq 4, \text{ for } k > 1$$

more referees[2]. Therefore, at least 6 referees are needed to cover all pairs of $n$ proposals in this case as well.

To complete the proof, suppose that $1 \leq u < k$. In this case, we must cover the pairs of proposals in all the sets stated above. In particular, we must cover the pairs of proposals in the sets $A \times C$, $A \times D$, $B \times D$, and $C \times D$. This leads to the assignment pattern for the subsequent referees as follows:

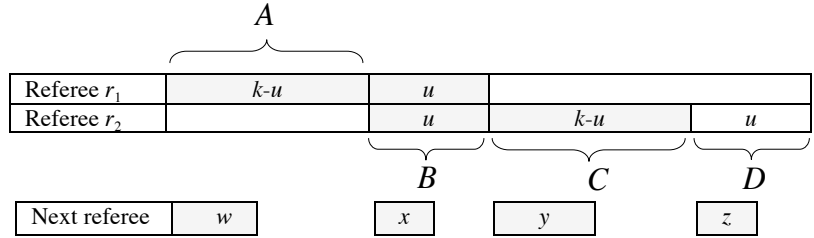

Figure 2.

Therefore, the number of additional referees cannot be less than

$$\left\lceil \frac{(k-u)(k-u+u) + (u+k-u)u}{wy + wz + xz + yz} \right\rceil = \left\lceil \frac{k^2}{wy + wz + xz + yz} \right\rceil$$

where $w$, $x$, $y$, $z$ are the numbers of proposals assigned to a new referee from the subsets, $A$, $B$, $C$, and $D$, respectively. It can be shown that, under the constraint $w + x + y + z = k$,

---

2  $\frac{3k^2-k}{k^2-k} > 3$ for all $k > 1$. Therefore, $\left\lceil \frac{3k^2-k}{k^2-k} \right\rceil \geq 4$ for all $k > 1$.

the denominator of the expression has a unique global maximum at $x = 0$, $w = y = z = k/3$, and is given by[3] $k^2/3$ (See Proposition 1(a) in the Appendix). However, since $u \geq 1$, the value of $x$ cannot be zero for all additional referees as this will leave out one or more pairs of proposals one of which belongs to $B$. Therefore, the maximum number of pairs generated by at least one of the additional referees must be less than $k^2/3$, and hence the number of additional referees cannot be less than 4. Adding these to the first two referees shows that 6 referees are necessary in this case as well and this completes the proof. ∥

**Corollary 1:** For all odd $n = 2k+1 \geq 5$, suppose that each of the half of the referees is assigned $k+1$ proposals, and each of the other half of the referees is assigned $k$ proposals. Then at least 6 referees are needed to cover all pairs of $n$ proposals.

**Proof:** Let $n = 2k+1$, where $k \geq 2$. Consider any $2k$ of the $n$ proposals, and let $p$ be the proposal that is left out. By Theorem 1, at least 6 referees must be used, with each assigned to $k$ proposals, to cover all $2k(2k-1)/2 = k(2k-1)$ pairs of these $2k$ proposals. This leaves

$$(2k+1)2k/2 - k(2k-1) = k\{(2k+1) - (2k-1)\} = 2k$$

pairs of proposals still to be covered. Suppose that one of the referees is removed and proposal $p$ is assigned to 3 of the remaining 5 referees each, in addition to their $k$ proposals which they had been originally assigned. Now, with one of the referees removed, at least one pair of proposals among the first $2k$ proposals, previously covered by the 6 referees must clearly be left uncovered. Otherwise, 5 referees would have been sufficient to cover the original $2k$ proposals. Therefore, at least $2k+1$ pairs of proposals must be covered by the three referees whose assignments have been increased by one proposal. However, with one new proposal, i.e., proposal $p$, these three referees can collectively increase the number of pairs of proposals by at most $2k$ since the three referees were assigned their $k$ proposals from the original set of $2k$ proposals prior to the assignment of proposal $p$. But, this is less than the $2k+1$ pairs of proposals still to be covered and the statement follows. ∥

These results can be extended to assignments where each referee can review $k = n/3$ proposals. For $n = 6$ ($k = 2$) and $n = 9$ ($k = 3$), it is easily verified that 15 and 12 referees are required. For all $n = 3k \geq 12$, where $k$ is a positive even integer, we can improve the lower bound of 10 referees in Table 1 to 11 as follows:

---

[3] This assumes that $k$ is divisible by 3. If it is not, the maximum becomes $(k^2-1)/3$ as mentioned in Proposition 1(a).

**Theorem 2:** For all $n = 3k \geq 12$, if each referee is assigned $k$ proposals then at least 11 referees are needed to cover all pairs of $n$ proposals[4].

**Proof:** The proof proceeds as in the proof of Theorem 1 with the following modified diagram. The only change in the set up is that the cardinality of $D$ is now $k+u$ and $k = n/3$.

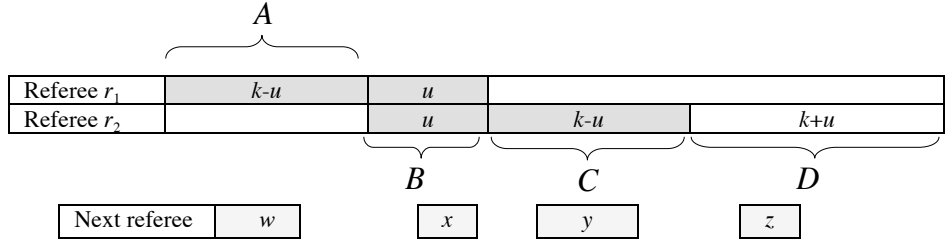

Figure 3.

As before, if $u = 0$ then $B$ vanishes, $A$, $C$, and $D$ contain $k$ proposals each without any overlap with one another. Hence, the number of pairs of proposals that remains to be covered is given by $3k^2 + k(k-1)/2$. Just considering the first term, the number of additional referees cannot be less than

$$\left\lceil \frac{3k^2}{wy + wz + yz} \right\rceil,$$

where $0 < w, y, z < k$ are the numbers of proposals in sets $A$, $C$, and $D$, and $w + y + z = k$. With the help of Proposition 1(a), it can be shown that the maximum value of $wy + wz + yz$ does not exceed $k^2/3$. Therefore, the minimum value of the expression above is given by

$$\left\lceil \frac{3k^2}{\frac{k^2}{3}} \right\rceil = 9.$$

proving that the total number of referees cannot be less than 11 in this case.

On the other hand, if $u = k$ then $A$ and $C$ vanish, and $|B| = k$, $|D| = 2k$ so that the number of proposals that remains to be covered is given by $2k^2 + 2k(2k-1)/2 = 4k^2 - k$. But since each new referee can cover at most $k(k-1)/2$ proposals, we need at least

$$\left\lceil \frac{4k^2 - k}{k(k-1)/2} \right\rceil = \left\lceil \frac{8k^2 - 2k}{k^2 - k} \right\rceil \geq 9$$

additional referees[5]. Adding these to the first two referees gives at least 11 referees to cover all pairs of $n$ proposals in this case as well.

---

[4] The statement can be extended to $n = 3k-1$ and $n = 3k-2$ using a similar argument as in Corollary 1.

Finally, suppose that $1 \leq u < k$. As in Theorem 1, we must cover the pairs of proposals in all the sets described in Eqn. (3). In particular, the number of pairs in the first four sets must be covered, where $A$, $B$, $C$, and $D$ are defined as in Figure 3. The number of these pairs of proposals is given by

$$(k-u)(k-u+k+u) + (u+k-u)(k+u) = 3k^2 - ku$$

With the distribution of $k$ proposals of each additional referee into the sets $A$, $B$, $C$, and $D$ as shown in Figure 3, the number of pairs of proposals covered by each additional referee is given by $wy + wz + xz + yz$. Furthermore, as shown in Proposition 1(b) in the Appendix, $wy + wz + xz + yz$ is maximized when $x = 0$, and $w = y$ for any given $z$. Therefore, the maximum number of pairs of proposals covered by each such referee is given by

$$wy + wz + xz + yz = w^2 + wz + wz = w^2 + 2wz$$

where $w + y + z = k$, or $z = k - w - y = k - 2w$. Replacing $z$ by $k-2w$ in the above equation, the maximum number of pairs of proposals that can be generated by any additional referee becomes

$$w^2 + 2w(k-2w) = 2kw - 3w^2$$

Let $a$ denote the minimum number of additional referees to cover the missing $3k^2$-$ku$ pairs of proposals, and let $w_i$ denote the number of proposals assigned to the $i$th referee, $1 \leq i \leq a$ under this maximality constraint. Then the maximum number of pairs of proposals covered by $a$ referees is given by

$$\sum_{i=1}^{a} 2kw_i - 3w_i^2.$$

Therefore, to cover the missing $3k^2$-$ku$ pairs, the following inequality must hold:

$$\sum_{i=1}^{a} 2kw_i - 3w_i^2 \geq 3k^2 - ku$$

Dividing both sides of the inequality by $k^2$ and rewriting the argument of the sum on the left, we get

$$\sum_{i=1}^{a} 2\frac{w_i}{k} - 3\frac{w_i^2}{k^2} = \sum_{i=1}^{a} \frac{w_i}{k}\left(2 - 3\frac{w_i}{k}\right) \geq 3 - \frac{u}{k}$$

It is easy to verify that the argument of the sum is maximized if

$$\frac{w_i}{k} = 2 - 3\frac{w_i}{k} \text{ or } w_i = \frac{k}{2}.$$

Therefore, the minimum value of $a$ satisfies the inequality

$$\sum_{i=1}^{a} 1/4 \geq 3 - \frac{u}{k} \text{ or } a \geq 12 - 4\frac{u}{k}.$$

---

[5] $\frac{8k^2-2k}{k^2-k} > 8$ for all $k > 1$. Therefore, $\left\lceil \frac{8k^2-2k}{k^2-k} \right\rceil \geq 9$ for all $k > 1$.

Given that
$$\left\lceil 12 - 4\frac{u}{k} \right\rceil \geq 9 \text{ if } u < k,$$
the number of additional referees cannot be less than 9, leading to a lower bound of 11 referees in this case as well. ∥

The next theorem extends these results to referees with a capacity of $n/4$ for $n$ proposals:

**Theorem 3:** For all $n = 4k \geq 16$, if each referee is assigned $k$ proposals, at least 18 referees must be used to cover all pairs of $n$ proposals.

**Proof:** Let $n = 4k$, where $k \geq 4$ is an integer. The proof proceeds as in the proofs of earlier theorems with the following modified diagram. The only change in the set up of the proof is that the cardinality of $D$ is now $2k+u$ and $k = n/4$.

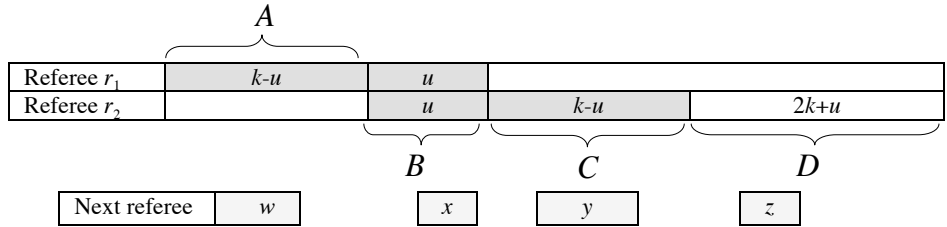

Figure 4.

If $u = 0$ then $B$ vanishes, $A$ and $C$ contain $k$ proposals each and $D$ contains $2k$ proposals without any overlap with one another. Hence, the number of proposals that remains to be covered is given by $5k^2 + k(2k-1)$. To cover the first $5k^2$ of these proposals, let $w$, $y$, $z$ be the number of proposals assigned to each additional referee from sets $A$, $C$, and $D$. Therefore, the number of additional referees cannot be less than

$$\left\lceil \frac{5k^2}{wy + wz + yz} \right\rceil$$

where $w + y + z = k$. As before, the maximum value of $wy + wz + yz$ cannot exceed $k^2/3$. Therefore, the minimum value of the expression above cannot be smaller than

$$\left\lceil \frac{5k^2}{\frac{k^2}{3}} \right\rceil = 15.$$

However, this assumes that the pairs of proposals generated by cross multiplying the sets of $k/3$ proposals from $A$, $C$, and $D$ can all be different. But, this is not possible since if we just consider the sets $A$ and $C$, and partition each into subsets of $k/3$ proposals then the maximum number of non-overlapping pairs of such subsets cannot exceed 9. Therefore, at least one pair of proposals must be covered more than once if we were to use more than 9 referees. This implies that the number of distinct pairs of proposals covered by cross multiplying subsets of $k/3$ proposals from each of *A, C*, and *D* must be less than $k^2/3$ for one or more of the additional referees. Therefore, at least 16 new referees are needed and adding this to the first two referees gives at least 18 referees.

On the other hand, if $u = k$ then *A* and *C* vanish, and $|B| = k$, $|D| = 3k$ so that the number of additional pairs of proposals to be covered is given by $3k^2+3k(3k-1)/2 = 15k^2/2-3k/2$. Dividing this by the maximum number of pairs of proposals that can be covered by a referee gives at least

$$\left\lceil \frac{15k^2/2 - 3k/2}{k(k-1)/2} \right\rceil = \left\lceil \frac{15k^2 - 3k}{k^2 - k} \right\rceil \geq 16$$

additional referees[6] or a total of 18 referees with the first two referees added.

Finally, suppose that $1 \leq u < k$. Given the distribution of the proposals to the sets *A, B, C*, and *D* as shown in Figure 4, the number of pairs of proposals that remain to be covered is given by

$$(k-u)3k + k(2k+u) + \frac{(2k+u)(2k+u-1)}{2} = 7k^2 - k + u^2 - u/2.$$

Now arbitrarily divide the set *D* into three subgroups of $D_1$, $D_2$ and $D_3$ where the sizes of these subgroups are $k$, $k$ and $u$ respectively and the sum of their sizes is $k + k + u = 2k + u$, the size of the set *D*. Suppose that the pairs of proposals within each of the sets $D_1$, $D_2$ and $D_3$ are already covered without using any new referees. Then the number of pairs of proposals that remain to be covered is given by

$$7k^2 - k + u^2 - u/2 - \left\{ \binom{k}{2} + \binom{k}{2} + \binom{u}{2} \right\} = 6k^2$$

Now, any additional referee can generate at most

$$wy + wz + ws + wt + xz + xs + xt + yz + ys + yt + zs + zt + st$$

new pairs of proposals as shown in figure below.

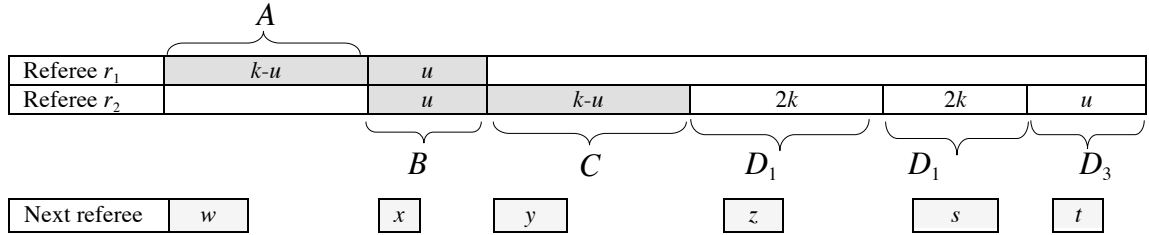

Figure 5.

Therefore the number of additional referees cannot be less than

$$\frac{6k^2}{wy + wz + ws + wt + xz + xs + yz + ys + yt + zs + zt + st}$$

---

[6] $\frac{15k^2-3k}{k^2-k} > 15$ for all $k > 1$. Therefore, $\left\lceil \frac{15k^2-3k}{k^2-k} \right\rceil \geq 16$ for all $k > 1$.

where $w$, $x$, $y$, $z$, $s$, and $t$ are the numbers of proposals assigned to a new referee from the subsets, $A$, $B$, $C$, $D_1$, $D_2$ and $D_3$, respectively. It can be shown that, under the constraint $w + x + y + z + s + t = k$, the denominator of this expression has a maximum at $w = y = z = s = t = k/5$ and $x = 0$, and is given by $2k^2/5$ (See Proposition 2). Hence the number of additional referees cannot be less than

$$\frac{6k^2}{2k^2/5} = 15.$$

However, this assumes that the pairs of proposals generated by cross multiplying the sets of $k/5$ proposals from $A$, $C$, $D_1$ $D_2$ and $D_3$ can all be different. But this is not possible since the number of non-overlapping pairs of subsets of size $k/5$ between $A$ and $D_3$ is strictly less than 15. To see this, just note that the number of non-overlapping pairs of subsets of size $k/5$ in $A$ is given by $(k-u)/(k/5)$ and similarly those in $D_3$ is given by $u/(k/5)$ Therefore, the maximum number of non-overlapping pairs of subsets of size $k/5$ is given by $(k-u)u/(k^2/25) = 25(k-u)u/k^2$. It is easy to see that $25(k-u)u/k^2$ is strictly less than 15 for any $u$, $1 < u \leq k$. It follows that the number of distinct pairs of proposals covered by cross multiplying subsets of $k/5$ proposals from each of $A$, $C$, $D_1$, $D_2$ and $D_3$ must be less than $2k^2/5$ for at least one of the additional referees. Hence the number of additional referees cannot be less than 16. Adding these to the first two referees shows that 18 referees are necessary in this case as well and this completes the proof. ∥

**3. Optimal Assignments With Indistinguishable Referees**

In this section, we provide explicit assignments of proposals to referees to cover all pairs of proposals using 6 referees for $n = 2k$, 12 referees for $n = 3k$, and 20 referees for $n = 4k$. We further prove that the lower bound of $\lceil n(n-1)/k(k-1) \rceil$ referees is asymptotically optimal within a factor of 2 by giving an actual assignment for capacity $k$ for all other $k$, $2 \leq k \leq n$.

**A. Indistinguishable Referees With Half and Some Other Fractional Capacities**

We first present an optimal assignment of $n$ proposals to referees with a capacity of $n/2$.

**Theorem 4:**

(a) For any even integer $n = 2k \geq 4$, if 4 referees are assigned $k$ proposals each, one referee is assigned $2\lceil k/2 \rceil$ proposals and one referee is assigned $2\lfloor k/2 \rfloor$ proposals, then 6 referees are sufficient to cover all pairs of $n$ proposals.

(b) For any odd integer $n = 2k+1 \geq 5$, if one half of referees are assigned $\lceil n/2 \rceil$ proposals and the other half of referees are assigned $\lfloor n/2 \rfloor$ proposals then 6 referees are sufficient to cover all pairs of $n$ proposals.

**Proof:**

(a) For even *n*, we give one possible assignment that uses 6 referees below.

| Proposals→ | $p_1, ..., p_{\lceil k/2 \rceil}$ | $p_{\lceil k/2 \rceil+1}, ..., p_k$ | $p_{k+1}, ..., p_{k+\lceil k/2 \rceil}$ | $p_{k+\lceil k/2 \rceil+1}, ..., p_{2k}$ |
|---|---|---|---|---|
| Referee $r_1$ | *k* proposals | | | |
| Referee $r_2$ | | | *k* proposals | |
| Referee $r_3$ | $\lceil k/2 \rceil$ proposals | | $\lceil k/2 \rceil$ proposals | |
| Referee $r_4$ | $\lceil k/2 \rceil$ proposals | | | $\lfloor k/2 \rfloor$ proposals |
| Referee $r_5$ | | $\lfloor k/2 \rfloor$ proposals | $\lceil k/2 \rceil$ proposals | |
| Referee $r_6$ | | $\lfloor k/2 \rfloor$ proposals | | $\lfloor k/2 \rfloor$ proposals |

Table 2. Assignment of *n* = 2*k* proposals to 6 referees, each with a capacity of *k*.

That this assignment covers all *n*(*n*-1)/2 pairs of proposals can be seen as follows. The first referee covers the *k*(*k*-1)/2 pairs of the first *k* proposals and the second referee covers the *k*(*k*-1)/2 pairs of the second *k* proposals, and therefore they are disjoint. The third referee covers $\lceil k/2 \rceil \times \lceil k/2 \rceil$ pairs of proposals and clearly, these pairs are all different from those covered by the first two referees. Likewise, the fourth, fifth, and sixth referees, cover $\lceil k/2 \rceil \times \lfloor k/2 \rfloor$, $\lfloor k/2 \rfloor \times \lceil k/2 \rceil$, $\lfloor k/2 \rfloor \times \lfloor k/2 \rfloor$ pairs of proposals which are all distinct from one another and those covered by the first three referees. Hence, the number of pairs covered by the 6 referees is given by

$$2k(k-1)/2 + \left\lceil \frac{k}{2} \right\rceil \times \left\lceil \frac{k}{2} \right\rceil + \left\lceil \frac{k}{2} \right\rceil \times \left\lfloor \frac{k}{2} \right\rfloor + \left\lfloor \frac{k}{2} \right\rfloor \times \left\lceil \frac{k}{2} \right\rceil + \left\lfloor \frac{k}{2} \right\rfloor \times \left\lfloor \frac{k}{2} \right\rfloor$$

$$= k(k-1) + \left\lceil \frac{k}{2} \right\rceil \times \left\{ \left\lceil \frac{k}{2} \right\rceil + \left\lfloor \frac{k}{2} \right\rfloor \right\} + \left\lfloor \frac{k}{2} \right\rfloor \times \left\{ \left\lceil \frac{k}{2} \right\rceil + \left\lfloor \frac{k}{2} \right\rfloor \right\}$$

$$= k(k-1) + \left\lceil \frac{k}{2} \right\rceil \times k + \left\lfloor \frac{k}{2} \right\rfloor \times k$$

$$= k(k-1) + \left\{ \left\lceil \frac{k}{2} \right\rceil + \left\lfloor \frac{k}{2} \right\rfloor \right\} \times k = k(k-1) + k^2 = \binom{2k}{2} = \binom{n}{2}$$

as desired.

(b) For odd *n* = 2*k*+1, we give the following assignment that also uses 6 referees.

| Proposals→ | $p_1, ..., p_{\lceil (k+1)/2 \rceil}$ | $p_{\lceil (k+1)/2 \rceil+1}, ..., p_{k+1}$ | $p_{k+2}, ..., p_{k+1+\lceil k/2 \rceil}$ | $p_{k+2+\lceil k/2 \rceil}, ..., p_{2k+1}$ |
|---|---|---|---|---|
| Referee $r_1$ | *k*+1 proposals | | | |
| Referee $r_2$ | | | *k* proposals | |
| Referee $r_3$ | $\lceil (k+1)/2 \rceil$ proposals | | $\lceil k/2 \rceil$ proposals | |
| Referee $r_4$ | $\lceil (k+1)/2 \rceil$ proposals | | | $\lfloor k/2 \rfloor$ proposals |
| Referee $r_5$ | | $\lfloor (k+1)/2) \rfloor$ proposals | $\lceil k/2 \rceil$ proposals | |
| Referee $r_6$ | | $\lfloor (k+1)/2) \rfloor$ proposals | | $\lfloor k/2 \rfloor$ proposals |

Table 3. Assignment of *n* proposals to 6 referees, each with a capacity of *n*/2, *n* = 2*k*+1.

As before, adding all the pairs of proposals contributed by the 6 referees, we obtain

$$\frac{(k+1)k}{2} + \frac{k(k-1)}{2} + \left\lceil\frac{k+1}{2}\right\rceil \times \left\lceil\frac{k}{2}\right\rceil + \left\lceil\frac{k+1}{2}\right\rceil \times \left\lfloor\frac{k}{2}\right\rfloor + \left\lfloor\frac{k+1}{2}\right\rfloor \times \left\lceil\frac{k}{2}\right\rceil + \left\lfloor\frac{k+1}{2}\right\rfloor \times \left\lfloor\frac{k}{2}\right\rfloor$$

$$= k^2 + \left\lceil\frac{k+1}{2}\right\rceil \times \left\{\left\lceil\frac{k}{2}\right\rceil + \left\lfloor\frac{k}{2}\right\rfloor\right\} + \left\lfloor\frac{k+1}{2}\right\rfloor \times \left\{\left\lceil\frac{k}{2}\right\rceil + \left\lfloor\frac{k}{2}\right\rfloor\right\}$$

$$= k^2 + \left\lceil\frac{k+1}{2}\right\rceil \times k + \left\lfloor\frac{k+1}{2}\right\rfloor \times k$$

$$= k^2 + \left\{\left\lceil\frac{k+1}{2}\right\rceil + \left\lfloor\frac{k+1}{2}\right\rfloor\right\} \times k = k^2 + (k+1)k = \binom{2k+1}{2} = \binom{n}{2}$$

and the statement follows. ∥

**Example 1:** (a) $n = 6, k = 3$      (b) $n = 8, k = 4$

| | | | | | | | | | | | | | | |
|---|---|---|---|---|---|---|---|---|---|---|---|---|---|---|
| $r_1$ | $p_1$ | $p_2$ | $p_3$ | | | | $r_1$ | $p_1$ | $p_2$ | $p_3$ | $p_4$ | | | |
| $r_2$ | | | | $p_4$ | $p_5$ | $p_6$ | $r_2$ | | | | | $p_5$ | $p_6$ | $p_7$ | $p_8$ |
| $r_3$ | $p_1$ | $p_2$ | | $p_4$ | $p_5$ | | $r_3$ | $p_1$ | $p_2$ | | | $p_5$ | $p_6$ | |
| $r_4$ | $p_1$ | $p_2$ | | | | $p_6$ | $r_4$ | $p_1$ | $p_2$ | | | | | $p_7$ | $p_8$ |
| $r_5$ | | | $p_3$ | $p_4$ | $p_5$ | | $r_5$ | | | $p_3$ | $p_4$ | $p_5$ | $p_6$ | |
| $r_6$ | | | $p_3$ | | | $p_6$ | $r_6$ | | | $p_3$ | $p_4$ | | | $p_7$ | $p_8$ |

(c) $n = 5, k = 2$      (c) $n = 7, k = 3$

| | | | | | | | | | | | | | |
|---|---|---|---|---|---|---|---|---|---|---|---|---|---|
| $r_1$ | $p_1$ | $p_2$ | $p_3$ | | | $r_1$ | $p_1$ | $p_2$ | $p_3$ | $p_4$ | | | |
| $r_2$ | | | | $p_4$ | $p_5$ | $r_2$ | | | | | $p_5$ | $p_6$ | $p_7$ |
| $r_3$ | $p_1$ | $p_2$ | | $p_4$ | | $r_3$ | $p_1$ | $p_2$ | | | $p_5$ | $p_6$ | |
| $r_4$ | $p_1$ | $p_2$ | | | $p_5$ | $r_4$ | $p_1$ | $p_2$ | | | | | $p_7$ |
| $r_5$ | | | $p_3$ | $p_4$ | | $r_5$ | | | $p_3$ | $p_4$ | $p_5$ | $p_6$ | |
| $r_6$ | | | $p_3$ | | $p_5$ | $r_6$ | | | $p_3$ | $p_4$ | | | $p_7$ |

Table 4. Optimal assignments of proposals to referees for $n = 5,6,7,8$.

**Remark 1:** When $n$ and $k = n/2$ are both even, each referee is assigned exactly $k$ proposals in Theorem 4 and this conforms to the hypothesis of Theorem 1. However, when $k$ is odd, this happens only in an average sense. That is, the average number of proposals assigned to the 6 referees is still $k$ with one of the referees receiving $k+1$ proposals and another referee $k-1$ proposals as in (a) in the example above. We conjecture that it is impossible to cover all pairs of proposals if all 6 referees are assigned exactly $k$ proposals. For odd $n = 2k + 1$, the assignments of proposals to the 6 referees conforms to the hypothesis of Theorem 1 for both even and odd $k$ as can be seen in (c) and (d) in the example above. In particular, when $k$ is even, referees $r_3$ and $r_4$ are assigned $k+1$ proposals each and $r_5$ and $r_6$ are assigned $k$ proposals each. When $k$ is odd, referees $r_3$ and $r_5$ are assigned $k+1$ proposals each and $r_4$ and $r_6$ are assigned $k$ proposals each. ∥

We also note that the assignments of the proposals to the 6 referees in Theorem 4 are not unique. For even $n$, there exist $\binom{2k}{k}\binom{k}{k/2}\binom{k}{k/2}$ such assignments, where $\binom{2k}{k}$ represents the number of choices for the first two referees, and the last two terms represent the number of choices for the last four referees. Similarly, for odd $n$, there exist $\binom{2k+1}{k+1}\binom{k+1}{k/2}\binom{k}{k/2}$ such assignments. ∥

**Theorem 5:** Suppose that $n$ is divisible by 9, and let $k = n/3$. Then 12 referees are sufficient to cover all pairs of $n$ proposals.

**Proof:** Divide the set of $n$ proposals into three groups of $k$ proposals each, and use a different referee to review the $k$ proposals in each group. This covers $3k(k-1)/2$ pairs of proposals with 3 referees and
$$\binom{3k}{2} - 3k(k-1)/2 = 3k^2$$
pairs of proposals remain. Now, divide each group of $k$ proposals into 3 disjoint groups of $k/3$ proposals (see Table 5)
$$\mathcal{H}_i = \{G_{i,1}, G_{i,2}, G_{i,3}\}, i = 1,2,3$$
where $|G_{i,j}| = k/3$, $1 \leq i, j \leq 3$, and use 9 referees to cover the remaining pairs of proposals as follows: Assign the subsets of $k/3$ proposals across $\mathcal{H}_1$, $\mathcal{H}_2$, and $\mathcal{H}_3$ to 9 referees in such a way that (1) each referee is assigned exactly one subset of $k/3$ proposals in each $\mathcal{H}_i$, and (2) no two referees are assigned the same two subsets of $k/3$ proposals from any two different groups, $\mathcal{H}_i$ and $\mathcal{H}_j$, $1 \leq i \neq j \leq 3$. This ensures that the pairs of proposals covered by the referees will be distinct. Moreover, each referee is assigned exactly $k$ proposals, one subgroup of $k/3$ proposals from each of the inner 3 groups. That this can always be done is proved in Proposition 3 in the Appendix and illustrated in the example below. To complete the proof, it is sufficient to note that
$$3^2\binom{3}{2}\left(\frac{k}{3}\right)^2 = 3k^2$$
pairs of proposals are covered by the 9 referees, and adding it to the number of proposals covered by the first 3 referees gives
$$3k(k-1)/2 + 3k^2 = \binom{3k}{2} = \binom{n}{2}$$
pairs of proposals as desired. ∥

|              | $\mathcal{H}_1$ |     |     | $\mathcal{H}_2$ |     |     | $\mathcal{H}_3$ |     |     |
|--------------|-----|-----|-----|-----|-----|-----|-----|-----|-----|
| Referee $r_1$ | $k$ |     |     |     |     |     |     |     |     |
| Referee $r_2$ |     |     |     | $k$ |     |     |     |     |     |
| Referee $r_3$ |     |     |     |     |     |     | $k$ |     |     |
| Referee $r_4$ | $k/3$ | $k/3$ | $k/3$ | $k/3$ | $k/3$ | $k/3$ | $k/3$ | $k/3$ | $k/3$ |
| Referee $r_5$ | $k/3$ | $k/3$ | $k/3$ | $k/3$ | $k/3$ | $k/3$ | $k/3$ | $k/3$ | $k/3$ |
| Referee $r_{12}$ | $k/3$ | $k/3$ | $k/3$ | $k/3$ | $k/3$ | $k/3$ | $k/3$ | $k/3$ | $k/3$ |
| | $G_{1,1}$ | $G_{1,2}$ | $G_{1,3}$ | $G_{2,1}$ | $G_{2,2}$ | $G_{2,3}$ | $G_{3,1}$ | $G_{3,2}$ | $G_{3,3}$ |

(spanning header: $n = 3k$ proposals)

Table 5. Assignment of $n$ proposals to $3 + 9 = 12$ referees, each with capacity $k$.

**Example 2:** The assignment below covers all 153 pairs of 18 proposals with 12 referees with each referee assigned 6 proposals. Whether it is possible to use 11 referees to cover all 153 pairs remains an open problem. ‖

| | $p_1$ | $p_2$ | $p_3$ | $p_4$ | $p_5$ | $p_6$ | $p_7$ | $p_8$ | $p_9$ | $p_{10}$ | $p_{11}$ | $p_{12}$ | $p_{13}$ | $p_{14}$ | $p_{15}$ | $p_{16}$ | $p_{17}$ | $p_{18}$ |
|---|---|---|---|---|---|---|---|---|---|---|---|---|---|---|---|---|---|---|
| $r_1$ | $p_1$ | $p_2$ | $p_3$ | $p_4$ | $p_5$ | $p_6$ | | | | | | | | | | | | |
| $r_2$ | | | | | | | $p_7$ | $p_8$ | $p_9$ | $p_{10}$ | $p_{11}$ | $p_{12}$ | | | | | | |
| $r_3$ | | | | | | | | | | | | | $p_{13}$ | $p_{14}$ | $p_{15}$ | $p_{16}$ | $p_{17}$ | $p_{18}$ |
| $r_4$ | $p_1$ | $p_2$ | | | | | $p_7$ | $p_8$ | | | | | $p_{13}$ | $p_{14}$ | | | | |
| $r_5$ | | | $p_3$ | $p_4$ | | | | | $p_9$ | $p_{10}$ | | | $p_{13}$ | $p_{14}$ | | | | |
| $r_6$ | | | | | $p_5$ | $p_6$ | | | | | $p_{11}$ | $p_{12}$ | $p_{13}$ | $p_{14}$ | | | | |
| $r_7$ | $p_1$ | $p_2$ | | | | | | | $p_9$ | $p_{10}$ | | | | | $p_{15}$ | $p_{16}$ | | |
| $r_8$ | | | $p_3$ | $p_4$ | | | | | | | $p_{11}$ | $p_{12}$ | | | $p_{15}$ | $p_{16}$ | | |
| $r_9$ | | | | | $p_5$ | $p_6$ | $p_7$ | $p_8$ | | | | | | | $p_{15}$ | $p_{16}$ | | |
| $r_{10}$ | $p_1$ | $p_2$ | | | | | | | | | $p_{11}$ | $p_{12}$ | | | | | $p_{17}$ | $p_{18}$ |
| $r_{11}$ | | | $p_3$ | $p_4$ | | | $p_7$ | $p_8$ | | | | | | | | | $p_{17}$ | $p_{18}$ |
| $r_{12}$ | | | | | $p_5$ | $p_6$ | | | $p_9$ | $p_{10}$ | | | | | | | $p_{17}$ | $p_{18}$ |

Table 6. Assignment of 18 proposals to 12 referees, each with a capacity of 6.

The previous two theorems can be extended to $n$ proposals and referees with capacity $n/4$.

**Theorem 6:** Suppose that $n$ is divisible by 16, and let $k = n/4$. Then 20 referees are sufficient to cover all pairs of $n$ proposals.

**Proof:** As in Theorem 5, divide the set of $n$ proposals into four groups of $k$ proposals each, and use a different referee to review the $k$ proposals in each group. This covers $4k(k-1)/2$ pairs of proposals with 4 referees and

$$\binom{4k}{2} - 4k(k-1)/2 = 6k^2$$

pairs of proposals remain. Now, divide each group of $k$ proposals into 4 disjoint groups of $k/4$ proposals

$$\mathcal{H}_i = \{G_{i,1}, G_{i,2}, G_{i,3}, G_{i,4}\}, i = 1,2,3,4$$

where $|G_{i,j}| = k/4$, $1 \leq i, j \leq 4$, and use 16 referees to cover the remaining pairs of proposals as follows: Assign the subsets of $k/4$ proposals across $\mathcal{H}_1$, $\mathcal{H}_2$, $\mathcal{H}_3$, and $\mathcal{H}_4$ to 16 referees in such a way that (1) each referee is assigned to exactly one subset of $k/4$ proposals in each $\mathcal{H}_i$, and (2) no two referees are assigned the same two subsets of $k/4$ proposals from any two different groups, $\mathcal{H}_i$ and $\mathcal{H}_j$, $1 \leq i \neq j \leq 4$ (See Proposition 4 in the Appendix). This ensures that the pairs of proposals covered by the referees will be distinct. Moreover, each referee is assigned exactly $k$ proposals, one subgroup of $k/4$ proposals from each of the inner 4 groups. To complete the proof, it is sufficient to note that

$$4^2 \binom{4}{2} \left(\frac{k}{4}\right)^2 = 6k^2$$

pairs of proposals are covered by the 16 referees, and adding it to the number of proposals covered by the first 4 referees gives

$$4k(k-1)/2 + 6k^2 = \binom{4k}{2} = \binom{n}{2}$$

pairs of proposals as desired. ∥

**B. Indistinguishable Referees With Arbitrary Capacity**

The assignments described in Theorems 4, 5, and 6 will work for effectively for small values of $n$. In particular, 6-referee assignments in Theorem 4 can handle up to 20 proposals where each referee may be assigned up to 10 proposals. However, for larger $n$, it will be impractical for referees to review $n/2$, $n/3$, or $n/4$ proposals and the number of proposals assigned to each referee may have to be decreased as needed. To deal with larger numbers of proposals, we present another assignment using an asymptotically minimum number of referees. The following theorem describes this assignment for any even $k$ that divides $n$. The theorem is easily extended to odd $k$ as described in the remark that follows the theorem.

**Theorem 7:** Let $n$ and $k$ be positive integers, where $k$ is even and divides $n$. It is sufficient to have $n(2n-k)/k^2$ referees, each with capacity $k$ to cover all $n(n-1)/2$ pairs of $n$ proposals.

**Proof:** Divide the set of $n$ proposals into $n/k$ groups, and use a different referee to review the $k$ proposals in each group. This covers $n\binom{k}{2}/k$ pairs with $n/k$ referees. Now, use four more referees to cover the pairs of proposals between every two groups of $k$ proposals as shown in Table 7 for one such pair of groups. This gives

$$4\binom{n/k}{2}\frac{k^2}{4} = \binom{n/k}{2}k^2$$

more distinct pairs, making the total number of pairs equal to

$$\frac{n}{k}\binom{k}{2} + \binom{n/k}{2}k^2 = \frac{n(k-1)}{2} + \frac{n(n-k)}{2} = \frac{n(n-1)}{2} = \binom{n}{2}$$

as desired. Since there are $\binom{n/k}{2}$ such pairs of groups, the number of referees we need to cover the pairs of proposals generated by these pairs of groups is given by $4\binom{n/k}{2}$. Therefore, the total number of referees to cover all $n(n-1)/2$ pairs of proposals is given by

$$\frac{n}{k} + 4\binom{n/k}{2} = \frac{n}{k} + 2\frac{n}{k}\left(\frac{n}{k}-1\right) = \frac{n(2n-k)}{k^2}$$

and the statement follows. ∥

**Corollary 2:** The number of referees used in the assignment described in Theorem 7 is within a factor of 2 of the lower bound given in Eqn. (2) and therefore is asymptotically optimal.

**Proof:** Dividing the number of referees obtained in Theorem 7 by the lower bound on the number of referees given in Eqn. (2), we get

$$\frac{n(2n-k)}{k^2} \times \frac{k(k-1)}{n(n-1)} = \frac{(2n-k)}{k} \times \frac{(k-1)}{(n-1)} < \frac{(2n-k)}{n-1} \leq 2, \text{ for } k \geq 2$$

and the statement follows. ∥

|  | $n$ proposals | | | | | | |
|---|---|---|---|---|---|---|---|
| Referee $r_1$ | $k$ | | | | | | |
| Referee $r_2$ | | $k$ | | | | | |
| Referee $r_3$ | | | $k$ | | | | |
| | | | | ... | | | |
| Referee $r_{n/k-1}$ | | | | | | $k$ | |
| Referee $r_{n/k}$ | | | | | | | $k$ |
| Referee $r_{n/k+1}$ | | | | $k/2$ | | $k/2$ | |
| Referee $r_{n/k+2}$ | | | | $k/2$ | | | $k/2$ |
| Referee $r_{n/k+3}$ | | | | | $k/2$ | $k/2$ | |
| Referee $r_{n/k+4}$ | | | | | $k/2$ | | $k/2$ |
| | | | | | | | |

Table 7. Assignment of $n$ proposals to $n(2n-k)/k^2$ referees, each with capacity $k$.

**Example 3** (Even $k$)**:** Let $n = 6$ and $k = 2$. By Theorem 7, $n(2n-k)/k^2 = 15$ referees are sufficient as illustrated in Table 8 below. In this case, the number of referees used does exactly match the minimum number of referees given in Eqn. (2).

| Referee $r_1$ | $p_1$ | $p_2$ | | | | |
| Referee $r_2$ | | | $p_3$ | $p_4$ | | |
| Referee $r_3$ | | | | | $p_5$ | $p_6$ |
| Referee $r_4$ | $p_1$ | | $p_3$ | | | |
| Referee $r_5$ | $p_1$ | | | $p_4$ | | |
| Referee $r_6$ | | $p_2$ | $p_3$ | | | |
| Referee $r_7$ | | $p_2$ | | $p_4$ | | |
| Referee $r_8$ | $p_1$ | | | | $p_5$ | |
| Referee $r_9$ | $p_1$ | | | | | $p_6$ |
| Referee $r_{10}$ | | $p_2$ | | | $p_5$ | |
| Referee $r_{11}$ | | $p_2$ | | | | $p_6$ |
| Referee $r_{12}$ | | | $p_3$ | | $p_5$ | |
| Referee $r_{13}$ | | | $p_3$ | | | $p_6$ |
| Referee $r_{14}$ | | | | $p_4$ | $p_5$ | |
| Referee $r_{15}$ | | | | $p_4$ | | $p_6$ |

Table 8. Assignment of $n = 6$ proposals to $n(2n-k)/k^2$ referees, each with capacity $k = 2$.

**Remark 2:** For odd $k$, partition the $n$ proposals into $n/k$ groups of $k$ proposals each as in Theorem 7 and assign each group to a different referee. Assign $k+1$ proposals to each of the rest of referees and divide each group of $k$ proposals into two overlapping groups of $(k+1)/2$ proposals as in the example below. The rest of the proof applies as it is.

| Referee $r_1$ | $p_1$ | $p_2$ | $p_3$ | | | |
| Referee $r_2$ | | | | $p_4$ | $p_5$ | $p_6$ |
| Referee $r_3$ | $p_1$ | $p_2$ | | $p_4$ | $p_5$ | |
| Referee $r_4$ | $p_1$ | $p_2$ | | | $p_5$ | $p_6$ |
| Referee $r_5$ | | $p_2$ | $p_3$ | $p_4$ | $p_5$ | |
| Referee $r_6$ | | $p_2$ | $p_3$ | | $p_5$ | $p_6$ |

(a) Assignment of 6 proposals to 6 referees.

| Referee $r_1$ | $p_1$ | $p_2$ | $p_3$ | | | |
| Referee $r_2$ | | | | $p_4$ | $p_5$ | $p_6$ |
| Referee $r_3$ | $p_1$ | | $p_3$ | $p_4$ | $p_5$ | |
| Referee $r_4$ | $p_1$ | $p_2$ | | | $p_5$ | $p_6$ |
| Referee $r_5$ | | $p_2$ | $p_3$ | $p_4$ | | $p_6$ |

(b) Assignment of 6 proposals to 5 referees.

Table 9.

**Example 4** (Odd $k$): Let $n = 6$ and $k = 3$. By Eqn. (2), 5 referees are necessary and by Theorem 7, $n(2n-k)/k^2 = 6$ referees are sufficient as shown in Table 9(a). As seen in the table, the proposals assigned to referees $r_3$, $r_4$, $r_5$, and $r_6$ overlap. This results in some of the pairs of proposals to be covered more than once but it does not increase the number of referees in the assignment. However, it also makes the assignment asymmetric with respect to the number of referees assigned to the proposals (proposals $p_2$ and $p_5$ are reviewed by 5 referees whereas the rest of proposals are reviewed by 3 referees each). This can be avoided by removing the last referee, and reassigning the proposals to remaining referees as shown in Table 9(b). ∥

## 4. Assignments With Distinguishable Referees

In the assignment problems considered thus far we have not taken into account the specialties of referees in handling proposals. It is often desirable to assign proposals to referees who are experts or specialists on the subjects of proposals they review. The assignment methods in Section 3 can still be applied if the specialties of referees satisfy certain constraints. In what follows, we describe some of these extensions.

**Corollary 3:** Suppose that a set of $n$ proposals can be partitioned into two specialty areas of $n/2$ proposals, $S_1$ and $S_2$. Further suppose that, among some 6 referees, (a) one is able to review the proposals in $S_1$ and another is able to review the proposals in $S_2$, and (b) the other four are each able to review $n/4$ proposals in each of $S_1$ and $S_2$. Then all pairs of $n$ proposals can be covered by the 6 referees with the side condition that each proposal is reviewed by three referees in its subject area.

**Proof:** It follows directly from Theorem 4 as shown in Table 10. ∥

| Proposals→ | $p_1, p_2, \ldots, p_{n/4}$ | $p_{n/4+1}, p_{n/4+2}, \ldots, p_{n/2}$ | $p_{n/2+1}, p_{n/2+2}, \ldots, p_{3n/4}$ | $p_{3n/4+1}, p_{3n/4+2}, \ldots, p_n$ |
|---|---|---|---|---|
| Referee 1 | Specialty $S_1$ | Specialty $S_1$ | | |
| Referee 2 | | | Specialty $S_2$ | Specialty $S_2$ |
| Referee 3 | Specialty $S_1$ | | Specialty $S_2$ | |
| Referee 4 | Specialty $S_1$ | | | Specialty $S_2$ |
| Referee 5 | | Specialty $S_1$ | Specialty $S_2$ | |
| Referee 6 | | Specialty $S_1$ | | Specialty $S_2$ |

Table 10. Assignment of proposals to 6 referees with 2 specialties.

This corollary can be generalized to $n/k$ specialty areas of $k$ proposals and $n/k + 4\binom{n/k}{2}$ referees for any integer $k$, $2 \leq k \leq n$ that divides $n$.

**Corollary 4:** Suppose that a set of $n$ proposals can be partitioned into $n/k$ subject areas of $k$ proposals, $S_1, S_2, \ldots, S_{n/k}$. Suppose also that (a) there exist $n/k$ referees with $n/k$ different specialties matching the $n/k$ subject areas of these $n/k$ sets of $k$ proposals and each is able to review $k$ proposals, (b) the remaining $4\binom{n/k}{2}$ referees can be partitioned into $\binom{n/k}{2}$ groups of 4 referees so that the referees in each group have two specialties matching the specialties of a distinct pair of sets of $k$ proposals. Then $n/k + 4\binom{n/k}{2}$ referees are sufficient to cover all pairs of $n$ proposals.

**Proof:** The proof immediately follows from Theorem 7. ∥

The example below illustrates the corollary.

**Example 5:** $n = 12$, $k = 4$. A possible assignment is shown in Table 11 with the following referee specialties:

Referee 1:   Specialty in $p_4, p_8, p_2, p_{11}$
Referee 2:   Specialty in $p_7, p_3, p_5, p_{10}$
Referee 3:   Specialty in $p_6, p_1, p_{12}, p_9$
Referee 4:   Specialty in $p_4, p_8, p_2, p_{11}$ and in $p_7, p_3, p_5, p_{10}$
Referee 5:   Specialty in $p_4, p_8, p_2, p_{11}$ and in $p_7, p_3, p_5, p_{10}$
Referee 6:   Specialty in $p_4, p_8, p_2, p_{11}$ and in $p_7, p_3, p_5, p_{10}$
Referee 7:   Specialty in $p_4, p_8, p_2, p_{11}$ and in $p_7, p_3, p_5, p_{10}$
Referee 8:   Specialty in $p_4, p_8, p_2, p_{11}$ and in $p_6, p_1, p_{12}, p_9$
Referee 9:   Specialty in $p_4, p_8, p_2, p_{11}$ and in $p_6, p_1, p_{12}, p_9$
Referee 10: Specialty in $p_4, p_8, p_2, p_{11}$ and in $p_6, p_1, p_{12}, p_9$
Referee 11: Specialty in $p_4, p_8, p_2, p_{11}$ and in $p_6, p_1, p_{12}, p_9$
Referee 12: Specialty in $p_7, p_3, p_5, p_{10}$ and in $p_6, p_1, p_{12}, p_9$
Referee 13: Specialty in $p_7, p_3, p_5, p_{10}$ and in $p_6, p_1, p_{12}, p_9$
Referee 14: Specialty in $p_7, p_3, p_5, p_{10}$ and in $p_6, p_1, p_{12}, p_9$
Referee 15: Specialty in $p_7, p_3, p_5, p_{10}$ and in $p_6, p_1, p_{12}, p_9$

With $n = 12$, $k = 4$, Eqn. (2) gives a lower bound of 11 on the number of referees. This assignment uses four more referees than the minimum number of referees needed.

| | | | | |
|---|---|---|---|---|
| Referee 1  | $p_4, p_8, p_2, p_{11}$ | | | |
| Referee 2  | | $p_7, p_3, p_5, p_{10}$ | | |
| Referee 3  | | | | $p_6, p_1, p_{12}, p_9$ |
| Referee 4  | $p_4, p_8$ | | $p_7, p_3$ | |
| Referee 5  | $p_4, p_8$ | | $p_5, p_{10}$ | |
| Referee 6  | | $p_2, p_{11}$ | $p_7, p_3$ | |
| Referee 7  | | $p_2, p_{11}$ | $p_5, p_{10}$ | |
| Referee 8  | $p_4, p_8$ | | | $p_6, p_1$ |
| Referee 9  | $p_4, p_8$ | | | $p_{12}, p_9$ |
| Referee 10 | | $p_2, p_{11}$ | | $p_6, p_1$ |
| Referee 11 | | $p_2, p_{11}$ | | $p_{12}, p_9$ |
| Referee 12 | | | $p_7, p_3$ | $p_6, p_1$ |
| Referee 13 | | | $p_5, p_{10}$ | $p_6, p_1$ |
| Referee 14 | | | $p_7, p_3$ | $p_{12}, p_9$ |
| Referee 15 | | | $p_5, p_{10}$ | $p_{12}, p_9$ |

Table 11. Assignment of 12 proposals to 15 referees satisfying specialty constrains.

## 5. Concluding Remarks

We have explored the referee complexity of covering all pairs of $n$ proposals. A lower bound on the referee complexity of covering all pairs of $n$ proposals has been derived for any $n \geq 2$, and this lower bound has been strengthened for referee capacities, $n/2$, $n/3$, and $n/4$. Explicit assignments, which are asymptotically optimal with respect to the derived lower bounds, have been given for proposals with and without specialty classifications. Table 12 below lists the number of referees facilitated by these assignments and their simple extensions for typical panel sizes used in peer-review systems. The numbers in parentheses denote the minimum number of referees required by the lower bound in Eqn. (2) except for the cases when $k = n/2, n/3, n/4$, $n = 20$ and $k = 15$, and $n = 30$ and $k = 20$. The lower bounds for the former cases are derived from Theorems 1, 2, and 3. In the latter two cases, assigning two referees 15 (or 20) proposals each leaves 10 (or 20) proposals unpaired as illustrated below for $n = 15$.

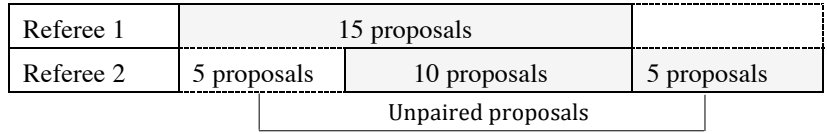

Figure 5.

Therefore, at least three referees are needed, and adding a third referee is sufficient to cover the missing pairs of proposals. The upper bounds are derived from Theorems 4, 5, 6, and 7. The shaded entries indicate the optimal assignments. The lower and upper bounds on the lower left are both unreasonably large and this is due to the fact that $k$ is very small compared to $n$. The upper bounds in this case are computed using Theorem 7 and both lower and upper bounds tend to $O(n^2)$ as $k$ tends to $O(1)$. On the other hand, as $k$ tends to $O(n)$, the lower and upper bounds both tend to $O(1)$. In particular, when $k = n/2$, the lower and upper bounds become 5 and 6, when $k = n/3$, they become 10 and 15, and when $k = n/4$, they become 17 and 28. Figure 6 depicts the lower and upper bounds based on the formulas $n(n-1)/k(k-1)$ and $n(2n-k)/k^2$ for $n = 50$ and $2 \leq k \leq 50$. It can be shown that the ratio of the upper bound to the lower bound reaches a maximum when $\sqrt{2n}$ for any $n$. It remains open if the lower and upper bounds can be made any closer, especially, for values of $k$ in the neighborhood of $\sqrt{2n}$.

Even though our results have been presented for assignments of proposals to referees, they can directly be applied to other assignment problems with similar constraints.

| Number of proposals ($n$) | Referee capacity ($k$) | | | |
|---|---|---|---|---|
| | 5 | 10 | 15 | 20 |
| 10 | 6(6) | 1(1) | 1(1) | 1(1) |
| 20 | 20(19) | 6(6) | 3(3) | 1(1) |
| 30 | 66(44) | 12(11) | 6(6) | 3(3) |
| 40 | 120(78) | 20(18) | 10(8) | 6(6) |
| 50 | 190(123) | 45(28) | 17(12) | 10(7) |

Table 12. Minimum and maximum numbers of referees to cover all pairs of proposals in typical proposal panels.

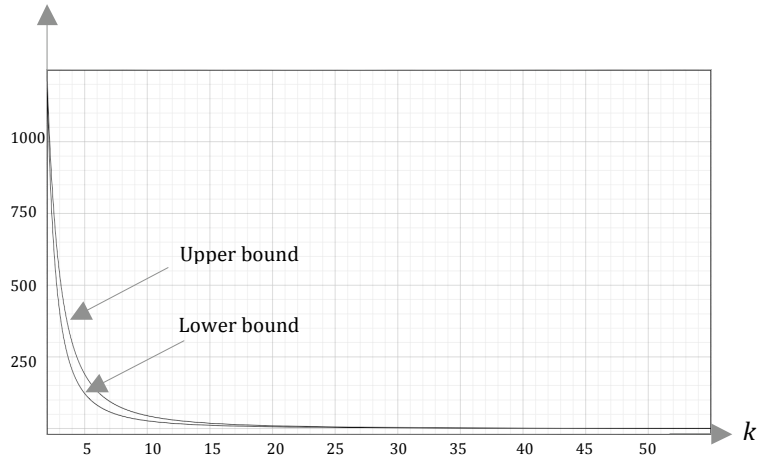

Figure 6. Lower and upper bounds for $n = 50$ and $2 \leq k \leq 50$.

## 6. Appendix

**Proposition 1:** Let $f(w,x,y,z) = wy + wz + xz + yz$.
  (a) Under the constraint $w + x + y + z = k$, the maximum value of $f(w,x,y,z)$ occurs at $x = 0$, $w = y = z = k/3$ and is at most equal to[7] $k^2/3$.
  (b) For any fixed $z$, and under the constraint $w + x + y + z = k$, the maximum value of $f(w,x,y,z)$ occurs when $w = y$, and $x = 0$.

**Proof:**

(a) Rearranging the terms in $f(w,x,y,z)$, we have $f(w,x,y,z) = wy + (w + x + y)z$ and since the second term $w + x + y$ can be increased arbitrarily by increasing $w$ and/or $y$ while also increasing the first term, setting $x = 0$ maximizes the value of $f(w,x,y,z)$. Now to find the maximum value of the function $f(w,0,y,z) = wy + wz + yz$ under the constraint $w + y + z = k$, it is sufficient to note that $f(w,0,y,z)$ is a symmetric function of $w$, $y$, and $z$, and therefore has its maximum when[7] $w = y = z = k/3$, and $f(k/3,0,k/3,k/3) = k^2/3$. Given that any value of $x$ other than 0 makes the product $wy$ less than $k^2/9$, at any global maximum of $f(w,x,y,z)$, $x$ must be 0. Similarly, since $f(w,0,y,z)$ is symmetric, any values of $w$, $y$, and $z$ other than $k/3$ should make $f(w,0,y,z)$ strictly less than $k^2/9$. Therefore, $f(w,x,y,z)$ has a unique maximum at $x = 0$, $w = y = z = k/3$.

(b) Using the same argument as in (a), for any $w$, $y$, and $z$, the maximum value of $f(w,x,y,z)$ must occur when $x = 0$. Then, for any fixed $z$, the constraint equation reduces to $w + y = k - z$. We can now determine the maximum value of $f(w,0,y,z)$ by setting up the Lagrangian,

$$L(w,y) = f(w,y) - \lambda(k - z - w - y)$$

---

[7] If $k$ is not evenly divisible by 3 then the maximum occurs at either $w=(k-1)/3+1$, $y=(k-1)/3$, $z=(k-1)/3$, or $w=(k-2)/3+1$, $y=(k-2)/3+1$, $z=(k-1)/3$ up to a permutation of $w$, $y$, and $z$. The direct substitution of $w$, $y$, and $z$ into $f(w,0,y,z)$ in each case shows that this maximum is $(k^2-1)/3$, and therefore, cannot exceed $k^2/3$.

and examining its derivatives with respect to $w$, $y$, and $\lambda$. This reveals that $f(w,0,y,z)$ assumes its maximum when $w = y = (k-z)/2$. ∥

**Proposition 2:** Let $f(w,x,y,z,s,t) = wy+wz+ws+wt+xz+xs+xt+yz+ys+yt+zs+zt+st$. Under the constraint $w + x + y + z + s + t = k$, the maximum value of $f(w,x,y,z,s,t)$ occurs at $x = 0$, $w = y = z = s = t = k/5$ and is equal to $2k^2/5$.

**Proof:** Rearranging the terms in $f(w,x,y,z,s,t)$, we have

$$f(w,x,y,z,s,t) = wy + (w+x+y)(z+s+t) + (s+t)z + st$$

and as in Proposition 1, setting $x = 0$ maximizes the value of $f(w,x,y,z,s,t)$. Now to find the maximum value of the function $f(w,0,y,z,s,t) = wy+wz+ws+wt+yz+ys+yt+zs+zt+st$ under the constraint $w + y + z + s + t = k$, it is sufficient to note that $f(w,0,y,z,s,t)$ is a symmetric function of $w$, $y$, $z$, $s$ and $t$, and therefore has its maximum when[8] $w = y = z = s = t = k/5$, and $f(k/5,0,k/5,k/5,k/5,k/5) = 2k^2/5$. ∥

**Lemma 1:** Let $U = \{1,2,3\}$, $V = \{4,5,6\}$, and $W = \{7,8,9\}$. There exists a set of nine triples $(u_i, v_i, w_i)$, $u_i \in U$, $v_i \in V$, and $w_i \in W$ such that the intersection of every two triples has at most one element in common.

**Proof:** The proof immediately follows from the following construction:

$$(1,4,7),(1,5,8),(1,6,9),(2,4,9),(2,5,7),(2,6,8),(3,4,8),(3,5,9),(3,6,7). \quad ∥$$

**Proposition 3:** As described in Theorem 5, suppose that each of the three groups of $k$ proposals is divided into three disjoint groups of $k/3$ proposals

$$\mathcal{H}_i = \{G_{i,1}, G_{i,2}, G_{i,3}\}, i = 1,2,3$$

where $|G_{i,j}| = k/3$, $1 \leq i, j \leq 3$. Further suppose that the subsets of $k/3$ proposals in $\mathcal{H}_1$, $\mathcal{H}_2$, and $\mathcal{H}_3$ are assigned to 9 referees in such a way that (1) each referee is assigned exactly one subset of $k/3$ proposals in each $\mathcal{H}_i$, and (2) no two referees are assigned the same two subsets of $k/3$ proposals from any two different groups, $\mathcal{H}_i$ and $\mathcal{H}_j$, $1 \leq i \neq j \leq 3$. Then the pairs of proposals generated by the 9 referees are all distinct.

**Proof:** The sets $\mathcal{H}_1$, $\mathcal{H}_2$, and $\mathcal{H}_3$ correspond to the sets $U$, $V$, and $W$ in Lemma 1, and their entries $G_{1,1}, G_{1,2}, G_{1,3}$; $G_{2,1}, G_{2,2}, G_{2,3}$; $G_{3,1}, G_{3,2}, G_{3,3}$ correspond to the entries in $U$, $V$, and $W$. Each of the triples

$$(1,4,7), (2,5,7), (3,6,7), (1,5,8), (2,6,8), (3,4,8), (1,6,9), (2,4,9), (3,5,9)$$

---

[8] If $k$ is not evenly divisible by 5, the maximum occurs at one of the following: $w = (k-1)/5+1$, $y = z = s = t = (k-1)/5$; $w = y = (k-2)/5+1$, $z = s = t = (k-2)/5$, $w = y = z = (k-3)/5+1$, $s = t = (k-3)/5$; $w = y = z = s = (k-4)/5+1$, $t = (k-4)/5$ up to a permutation of $w$, $y$, $z$, $s$, and $t$. Direct substitution of $w$, $y$, $z$, $s$, and $t$ into $f(w,0,y,z,s,t)$ in each case shows that the maximum values are $(2k^2-2)/5$, $(2k^2-3)/5$, $(2k^2-3)/5$, $(2k^2-2)/5$ respectively and therefore, cannot exceed $2k^2/5$.}

represents a Cartesian product of three entries from each of the sets $\mathcal{H}_1$, $\mathcal{H}_2$, and $\mathcal{H}_3$. For example, (1,4,7) represents the product

$$G_{1,1} \times G_{2,1} \times G_{3,1} = \{p_1, p_2\} \times \{p_7, p_8\} \times \{p_{13}, p_{14}\}$$

assuming that we use $n$ and $k$ given in Example 2. By Lemma 1, all these triples are distinct and no two have more than one element in common. Therefore, assigning a referee to each of the 9 triples ensures that all pairs of proposals generated by the 9 referees are distinct. ‖

**Lemma 2:** Let $U = \{1,2,3,4\}$, $V = \{5,6,7,8\}$, $W = \{9,10,11,12\}$, and $X = \{13,14,15,16\}$. There exists a set of sixteen quadruples $(u_i, v_i, w_i, x_i)$, $u_i \in U$, $v_i \in V$, $w_i \in W$, $x_i \in X$ such that the intersection of every two quadruples has at most one element in common.

**Proof:** The proof immediately follows from the following construction:

$$(1,5,9,13), (1,6,11,16), (1,7,12,14), (1,8,10,15),$$
$$(2,6,10,14), (2,5,12,15), (2,8,11,13), (2,7,9,16),$$
$$(3,7,11,15), (3,5,10,16), (3,6,12,13), (3,8,9,14)$$
$$(4,8,12,16), (4,5,11,14), (4,7,10,13), (4,6,9,15). \quad ‖$$

**Proposition 4:** As described in Theorem 6, suppose that each of the four groups of $k$ proposals is divided into four disjoint groups of $k/4$ proposals

$$\mathcal{H}_i = \{G_{i,1}, G_{i,2}, G_{i,3}, G_{i,4}\}, i = 1,2,3,4$$

where $|G_{i,j}| = k/4$, $1 \leq i, j \leq 4$. Further suppose that the subsets of $k/4$ proposals in $\mathcal{H}_1$, $\mathcal{H}_2$, $\mathcal{H}_3$, and $\mathcal{H}_4$ are assigned to 16 referees in such a way that (1) each referee is assigned exactly one subset of $k/4$ proposals in each $\mathcal{H}_i$, and (2) no two referees are assigned the same two subsets of $k/4$ proposals from any two different groups, $\mathcal{H}_i$ and $\mathcal{H}_j$, $1 \leq i \neq j \leq 4$. Then the pairs of proposals generated by the 16 referees are all distinct.

**Proof:** The proof is similar to the proof of Proposition 3 and omitted. ‖

*Last revised on August* 22, *2009.*